\documentclass[fleqn]{article}
\usepackage{graphics}

\usepackage{longtable}

\begin{document}

\newcommand{\keywords}{chemotaxis, brownian motion} 

\newcommand{\PACS}{87.10+e -- General theory and mathematical aspects }

\title{\"{U}ber Chemotaxis}
\author{S.\ L.\ Vesely$^{1}$} 

\newcommand{\address}
  {$^{1}$I.T.B., C.N.R., via Fratelli Cervi 93, I-20090 Segrate(MI)
   \\ \hspace*{0.5mm} Italy \\ 
   }

\newcommand{\email}{\tt sara.vesely@itb.cnr.it} 

\maketitle

{\small
\noindent \address
\par
\par
\noindent email: \email
}
\par

{\small
 \par
\par
\noindent{\bf PACS:} \PACS 
}

\begin{abstract}

Wir bringen eine Analogie zwischen der im Titel erw\"{a}hnten Eigenschaft 
von Lebewesen und den durch einen Temperaturgradienten hervorgerufenen 
Erscheinungen in Gasen. Dazu er\"{o}rtern wir zwei einfache mathematische 
Verfahren zur Behandlung von Messdaten.

\end{abstract}

\section{
Einleitung}

Chemotaxis ist ein bei Lebewesen beschriebener Vorgang \cite{adler1967}.
Darunter versteht man das Streben von ein- sowie mehrzelligen 
Wesen in Richtung auf den Ursprung eines chemischen Reizes oder
von ihm fort. Ersteres nennt man positive, zweiteres negative 
Chemotaxis. \\
Unser Interesse an dieser Erscheinung gilt haupts\"{a}chlich 
ihrer physikalischen Deutung. Im Unterschied zur rein mathematischen
Modellierung, bei der es sich um die Wiedergabe einer bereits
vorliegenden Erkl\"{a}rung handelt, sollte erstere
die unmittelbaren, vertrauten Erfahrungen an Lebewesen ansprechen,
und Kriterien und Mittel anwenden, die ihr eigen sind \cite{schroedinger1967}\cite{hilbert1992}\cite{bleecken1990}\cite{mazzag2002}. 
Damit w\"{u}nschen wir keineswegs einen Vergleich zwischen der 
physikalischen und der biologischen Auffassung anzustellen, sind
wir ja an keinem Wettbewerb um die Wahrheit, sondern an gemeinsame
Erkenntnisse beteiligt.\\
Eine ziemliche Verwirrung entsteht dennoch, wenn man beiderseits von dem
gleichen Bekenntnis ausgeht, dass biologische Abl\"{a}ufe aus den anorganischen
Verbindungen lauter au{\ss}erordentlichen und eigent\"{u}mlichen
Ereignissen zufolge auftauchen. Die Deutung eines Ph\"{a}nomens als
,,Wunder`` ist f\"{u}r sich genommen noch kein Widerspruch gegen seine
Erkl\"{a}rbarkeit. Weil sich jedoch Wundern eine gewisse Einmaligkeit
gesellt, und weil es sich gleichsam von selbst verbietet etwa vitalistische 
Prinzipien oder unwiederholbare Gesetzm\"{a}{\ss}igkeiten in die Physiklehre 
einzuf\"{u}hren, greift der Physiker schlie{\ss}lich zu reduktionistischen 
Ans\"{a}tzen.\\
Der angef\"{u}hrte Standpunkt - dass das Dasein alle \"{u}blichen 
Naturerscheinungen weit \"{u}bertrifft - ist, was die Erlebnisse 
des einzelnen Individuums betrifft, vollkommen gerechtfertigt, 
und auch vom biologischen und ethischen Gesichtspunkt haben wir f\"{u}r diese
Auffassung vieles \"{u}brig. Aber wenn es darum geht, ein nach physikalischen 
Kriterien vern\"{u}nftiges Bild wiederholbarer Ereignisse zu entwerfen, um 
bestimmte empirische und experimentelle Befunde zu deuten, erscheint 
uns zweckm\"{a}{\ss}iger die Lebewesen schlechthin zu den beschreibbaren 
Naturerscheinungen zu z\"{a}hlen. Dazu kommt, dass der Logos Leben 
eher begrifflich von fachwissenschaftlichen Bestimmungen gepr\"{a}gt 
als kausal durch das komplexe Zusammenwirken 
der Systembestandteile bedingt zu sein scheint.

Nachdem wir diesen uns am Herzen liegenden Punkt vorausgeschickt 
haben, m\"{o}chten wir noch zwei Punkte ber\"{u}hren. Zum ersten 
gibt es Tropismen, die wohl von Feldgradienten gesteuert werden 
k\"{o}nnten, darunter z.B. Phototropismus. In diesem Fall ruft das Lebewesen 
durch seine eigene Stellung im Lichtfeld eine St\"{o}rung hervor, 
die es aus der zur Quelle gewandten Seite als einen hellen Umriss, 
aus der gegen\"{u}berliegenden Seite als einen Schatten wahrnehmen 
l\"{a}sst. Im Unterschied dazu weist die Chemotaxis keinen starken 
Feldgradient auf. Wir sind nun der Meinung, dass eine physikalische 
Modellierung diesen Unterschied in der Beschreibung ber\"{u}cksichtigen 
sollte.\\
Der zweite Punkt ist folgender: Bei hochentwickelten Arten k\"{o}nnte 
man f\"{u}r positive Chemotaxis beispielsweise folgende Darstellungen 
bieten:
\begin{itemize}
\item
Das Tier nimmt das chemische Signal wahr;
\item
Es stellt die Windrichtung fest;
\item
Um das Ziel zu erreichen, bewegt es sich gegen 
den Wind.

\end{itemize}

Oder auch:
\begin{itemize}
\item
Das Tier bestimmt die Gr\"{o}{\ss}e des chemischen Signals;
\item

Es speichert sie und seine zeitweilige Lage;
\item

Es erreicht probeweise eine neue Stelle, oder \"{a}ndert 
nur seine Nasenspitzenstellung;
\item

Es sch\"{a}tzt die neue Intensit\"{a}t und vergleicht 
sie mit der gespeicherten Gr\"{o}{\ss}e;
\item

Es schl\"{a}gt die Richtung des zunehmenden Reizes 
ein.
\end{itemize}

Sollten sich derartige Darstellungen in allen F\"{a}llen als die 
einzig m\"{o}glichen erweisen, dann versch\"{o}be sich das Problem 
auf das Zentralnervensystem.\\
Doch wie steht es mit der Bewegung von Samenzellen zu den Eileitern? 
Sind Spermatozoen schon imstande etwas zu gewahren und sich daran 
zu erinnern? Wir werden versuchen diese wichtige Frage zu verneinen, 
indem wir eine von den Obengenannten abweichende Darstellung 
geben.

Die von uns zu beschreibenden Gegenst\"{a}nde sollen also, ohne 
das ortsabh\"{a}ngi-ge Konzentrationsgef\"{a}lle des Reizstoffes zu 
bemerken, sich in Richtung auf das chemische Ziel zu bewegen, 
und durch nichts weiteres als eine statistische
Geschwindigkeitsverteilung gekennzeichnet sein.

Als erstes m\"{o}chten wir ein Beispiel dieses statistischen Merkmals 
aus der Physiklehre entlehnen, damit jegliche Verkn\"{u}pfung mit 
der Biologie aus bleibt. Dann beschreiben wir zwei weitere mathematische 
Modelle, die uns einfach genug erscheinen, um dasselbe Merkmal 
deutlich erkennen zu lassen, und die in unserer Hoffnung einem 
experimentellen Forscher erlauben sollten, nach und nach das 
mathematische Verfahren mit den Versuchsergebnissen zu vergleichen.

\section{
Ankn\"{u}pfung an die kinetische Gastheorie}

Wir m\"{o}chten zun\"{a}chst das chemotaktische Verhalten der mit 
einem optischen Mikroskop gut beobachtbaren Einzeller der brownschen 
Bewegung weniger nicht absetzender K\"{o}rnchen in einer Wasserl\"{o}sung 
gegen\"{u}berstellen \cite{oppel1912}. Aus dem Vergleich zwischen
diesen beiden Erscheinungen ergibt sich letzten Endes das mathematische
Modell \cite{keller1971}\cite{boon1986}.  Beide Mal stellt sich w\"{a}hrend der Untersuchung 
ein Gef\"{a}lle ein. Ein Mal ist es ein von der Beleuchtung des Pr\"{a}parats
herr\"{u}hrendes Temperaturgef\"{a}lle zwischen der dem Kondensator
zugewandten Glasplatte des Objekttr\"{a}gers und dem Deckglas.
Das andere Mal haben die aufgel\"{o}sten Chemi-kalien auf den
Wanderungssinn der Zellen Einfluss. Wir werden hier
annehmen, dass der ganze Unterschied in der St\"{a}rke
dieser Steigung bestehe. Weil die mittlere freie Wegl\"{a}nge
der Zellen von ihrer Beweglichkeit, ihrer Konzentration und 
ihren Dimensionen abh\"{a}ngt, m\"{u}ssen die statistischen 
Schwankungen beider Systeme Verschiedenheiten aufweisen \cite{adler1967}. Aber 
der angenommenen \"{A}hnlichkeit mit der brownschen Bewegung wegen,
liefert dasselbe mathematische Modell unter g\"{u}nstigen Umst\"{a}nden,
d.h. wenn wir Konvektionsterme vernachl\"{a}ssigen d\"{u}rfen,
auch die mittlere Verschiebung der Einzeller.

Zur Anwendung der kinetischen Gastheorie auf das Ph\"{a}nomen der 
Chemotaxis, brauchen wir noch auf die thermischen Erscheinungen 
der Gase, die eben mittels der erw\"{a}hnten Theorie gedeutet wurden, 
Bezug zu nehmen. Dazu denken wir uns etwa die vertraute Erscheinung 
einer in feuchter Luft sich beschlagender Fensterscheibe. Wenn 
wir zwei eine Fl\"{u}ssigkeit enthaltende Reagenzgl\"{a}ser, von 
denen das eine auf einer tieferen Temperatur gehalten wird, mit 
einem leeren Rohr verbinden, beobachten wir, dass die Fl\"{u}ssigkeit 
sich im w\"{a}rmeren Probierglas verdampft und auf der k\"{a}lteren 
Glaswand verdichtet (und eventuell als Tr\"{o}pfchen herabrinnt). Bekanntlich 
erkl\"{a}rt die kinetische Theorie diesen Sachverhalt, indem sie 
sich vom Gas ein aus vielen Massenpunkten bestehendes Muster 
anfertigt. Danach verlassen die ,,Gasmolek\"{u}le`` statistisch 
den Ort, wo ihre vom W\"{a}rmezustand bedingte Geschwindigkeit 
die gr\"{o}{\ss}eren Werte hat, um zur kalten Wand hin zu wandern. 
Die gegen einen Reiz zu strebenden Einzeller bilden nun ein genaues 
Pendant dazu, wenn wir annehmen, dass ihre Bewegung in Richtung 
einer Verringerung der Geschwindigkeit erfolgt. Unserer 
Beschreibung nach wirkt folglich der chemische Reiz bei positiver 
Chemotaxis als Hemmstoff der Geschwindigkeit, w\"{a}hrend umgekehrt 
im Fall negativer Chemotaxis die Geschwindigkeit mit der Entfernung 
von der Quelle zunehmen soll. Nun heben wir hervor, dass das 
\"{u}bernommene Modell im Prinzip direkt auf die Daten Anwendung 
findet, da die relative Lage der Zellen anhand des Mikroskops 
tats\"{a}chlich ermittelt werden kann. Deshalb st\"{u}tzt sich weder 
die positive noch die negative Chemotaxis auf thermodynamische 
Prinzipien und beide modellieren weiterhin zwei voneinander unabh\"{a}ngige 
Erscheinungen. Es handelt sich um ein Modell, weil Einzeller 
nicht in jeder Hinsicht als Punkte darstellbar sind.

Um das Molek\"{u}lmodell durch ein f\"{u}r Zellen passendes Modell 
zu ersetzen, brauchen wir also in der Zustandsgleichung 
idealer Gase den Mittelwert der Geschwindigkeit der Zellen anstelle 
der Temperatur, und deren Konzentration mal derselben Geschwindigkeit 
statt des Druckes anzuwenden.\\
Wir m\"{o}chten zum Schluss folgende zwei Bemerkungen machen:\\
\begin{itemize}
\item
Zellen sind Energieverbraucher. Erhaltungss\"{a}tze f\"{u}r mechanische Energie und Impuls 
gelten dagegen f\"{u}r den vollkommen elastischen Stoss zwischen 
Gasmolekeln.\\
\item
Die Proportionalit\"{a}t des Wertes von Konzentration mal Temperatur zum Druck 
entspricht keiner Eigenschaft des Systems mehr. 
Sie ist eher als Ansatz anzusehen.
\end{itemize}

\section{
Anwendung von Differentialgleichungen}

Die Auffassung, wonach die Chemotaxis von einem \"{u}ber der Umgebung 
des Reizes aufgebauten (stetigen) Konzentrationsgradient eines 
chemischen Stoffes verursacht wird, scheint uns nicht ganz stichhaltig. 
Zum ersten: Physikalisch spielt dieser Gradient, was das Diffusionsbestreben 
der Zellen betrifft, etwa die Rolle eines osmotischen Drucks. Da laut Annahme 
die Erscheinung auf  Chemikalien beruht, muss 
sich die \"{o}rtliche Funktionalabh\"{a}ngigkeit des jeweiligen Stoffes 
angeben lassen. Aber entweder ist die Punktkonzentration des 
Reizes \"{u}berall verschwindend klein, oder dessen Dichte, als 
Masse pro Volumeneinheit definiert, w\"{a}chst \"{u}ber alle Ma{\ss}en. 
Beruft man sich statt dessen auf eine mittlere Belegungsdichte, 
dann treten die Schwierigkeiten bei der Erkl\"{a}rung des Konzentrationsgradienten 
auf \cite{einstein1908}.\\
Zweitens. Soll die (stetige) Fortbewegung eines Einzellers
strikt von der lagebedingten Versorgung mit Nahrungsmitteln 
bzw. Giften gesteuert werden, dann muss irgendein Mechanismus sowohl 
die Feststellung ihrer Zunahme als auch die \"{U}bertragung 
des Verspeisten auf Bewegung erlauben. In erster Ann\"{a}herung, 
d.h. wenn das Tierchen als punktf\"{o}rmig angenommen wird, scheint 
uns die Vorstellung, wonach zwei angebrachte F\"{u}hler das Konzentrationsunterschied 
zu ermitteln erlaubten, nicht ganz befriedigend. Au{\ss}erdem kann 
die Frage, ob der Zuwachs nicht zu geringf\"{u}gig und zu sehr 
mit Schwankungen behaftet sei, um eine gerichtete Fortbewegung 
im Gang zu setzen, trotzdem aufgeworfen werden. Wird der chemische Reiz \"{u}ber
Chemorezeptoren ausschliesslich als Signal gespeichert, dann kommt die 
Umwandlung in Bewegung der Einschaltung eines Zentralnervensystems 
gleich \cite{macnab1972}\cite{armstrong1967}.\\
Drittens. Wenn wir auf das mechanische Verst\"{a}ndnis des biologisch 
ohnehin komplizierten Vorgangs \cite{berg1986}\cite{viggiano1979} verzichten und die positive Chemotaxis 
laut der im vorigen Paragraph vorgeschlagenen \"{A}hnlichkeit mit einer 
Verz\"{o}gerung in Zusammenhang bringen, d\"{u}rfen wir uns nicht 
mehr auf das gew\"{o}hnte Newtonsche Bild der Bewegung st\"{u}tzen. 
Entsprechend erfahren Raum- und Zeitkoordinaten keine dynamische 
Deutung mehr, weshalb wir nur mangels eines besseren Ausdruckes 
von einer Diffusions\textit{geschwindigkeit} weiterreden. Die mittlere 
Einzellerlage und deren Zeitabh\"{a}ngige Verteilung k\"{o}nnen aber 
in dem Wahrscheinlichkeitsraum gedeutet werden.

Wenn die Konzentration des die Beschleunigung oder die Verz\"{o}gerung 
erregenden Stoffes station\"{a}r und von der Zellenbewegung unabh\"{a}ngig 
ist, dann h\"{a}ngt der Betrag der Zellengeschwindigkeit nur vom Ort
(Zufallsvariable) ab. In einem eindimensionalen Bezugssystem (z.B. einem 
Rohr mit konstantem Durchmesser), gilt dann folgende Beziehung:

\begin{equation}
{\partial\over\partial t}{c(x,t)}\propto {\partial^2\over\partial x^2}\bigl[c(x,t)|\nu(x)|\bigr]
\end{equation}

bei deren Ableitung man nach \"{a}hnlichen Methoden wie bei Diffusionsgleichungen 
vorgeht, wobei der Geschwindigkeitsbetrag 
$\left| v(x)\right| $
 ausdr\"{u}cklich eingef\"{u}hrt worden ist.
$c(x.t)$
 bezeichnet die Zellenanzahl pro Volumeneinheit.\\
Im Folgenden nehmen wir an, dass die Zellen sich auf einer im Ursprung des Achsenkreuzes
liegenden Quelle zusammendr\"{a}ngen  (positive Chemotaxis). 
Wir begn\"{u}gen uns mit der Anwendung auf den station\"{a}ren Zustand, also wenn:

\begin{equation}
{\partial\over\partial t}{c(x,t)}=0
\end{equation}

gilt, so dass, wenn wir:

\begin{equation}
{c(x,t)}=c_\infty(x)
\end{equation}

setzen, die neue Gleichung

\begin{equation}
{\partial^2\over\partial x^2}\bigl[c_\infty(x)|\nu(x)|\bigr]=0
\end{equation}

gilt, deren L\"{o}sung

\begin{equation}
c_\infty(x)|\nu(x)|=k_1 x+k
$$
\end{equation}

lautet. Da aber f\"{u}r ein willk\"{u}rliches $x$
$c_{\infty }^{} (x)\geq 0$
 und 
$\left| v(x)\right| \geq 0$
 gelten sollen, muss k$_{1}$ = 0 sein. Somit erh\"{a}lt die Gleichung 
die von uns anzuwendende Form:

\begin{equation}
c_\infty(x)={k \over {|\nu(x)|}}
\end{equation}

Da die Geschwindigkeit ein Minimum f\"{u}r $x$ = 0, wo 
sich die Quelle befindet, besitzt, hat hier 
$c_{\infty } (x)$
 sein Maximum, wenn die (6) bez\"{u}glich Null beschr\"{a}nkt 
bleibt.\\
Als Anwendung auf unseres Modell f\"{u}hren wir noch zwei gegens\"{a}tzliche 
M\"{o}glich-keiten an.

\subsection{
Positive Chemotaxis, 1. Grenzverteilung}

Wenn der Hemmstoff ausreichend schwach ist, um
$\left| v(x)\right| >0$
 auch bei Null zu gen\"{u}gen, und der Raum sich zwischen den Grenzen
 $\left[-1,+1\right]$
 erstreckt, dann ist
$c_{\infty } (x)$
 beschr\"{a}nkt, und man erkennt:

$\int\limits_{-1}^{+1}c_{\infty }  (x)=N_{tot} $

 und

$k=\frac{N_{tot} }{\int\limits_{-1}^{+1}\frac{dx}{v(x)}  } $

 wo N$_{tot}$ die Zellenanzahl ist.

\subsection{
Positive Chemotaxis, 2. Grenzverteilung}

Wenn der Inhibitor stark genug ist, folgende Bedingungen 
zu erf\"{u}llen:

$\left| v(x)=0\right| $

in $x$ = 0, und 

$\int\limits_{-1}^{+1}\frac{dx}{v(x)}  \rightarrow \infty $ ,

dann h\"{a}ufen sich die Zellen im Ursprung auf.

\section{
Anwendung von Differenzengleichungen}

Wenn wir eine geringe Anzahl von Einzellern im Mikroskop betrachten und 
die Beobachtungen nur f\"{u}r eine endliche Zeitspanne vornehmen, 
k\"{o}nnen wir zur Darstellung der Zellenfortpflanzung
statt der Differentialgleichungen Differenzen anwenden.\\
Fassen wir zun\"{a}chst einen einzigen Einzeller ins Auge \cite{berg1972}. Wenn 
wir jeweils dessen Verlagerung zum vorgegebenen Zeitpunkt notierten, 
handelten wir als ob wir bestrebt w\"{a}ren den zeitlichen Ablauf 
der Bewegung eines Massenpunktes in tabellarischer Form zusammenzufassen. 
Wird die Tabelle f\"{u}r  \"{a}quidistante Zeiten angeschrieben, 
wie wir jetzt voraussetzen wollen, so sind die Differenzen (erste 
Steigungen) bereits in der von uns anzuwendenden Form erkl\"{a}rt. 
Aus diesen Steigungen l\"{a}sst sich hinterher der Zeitverlauf 
anhand der Newtonschen Interpolationsformel berechnen. Obwohl 
sich ein bestimmter Verlauf ergibt, \"{a}ndert jede neue Messung, 
sowie jede Wahl einer gr\"{o}beren Zeitfolge auf die bereits angesammelten 
Daten das Ergebnis stark. Rein analytisch besagt das nur, dass 
der Steigungen keine eindeutige Kurve zugeordnet ist denn, gesetzt 
die Kurve w\"{a}re differenzierbar, w\"{u}rde die Interpolationsformel 
zur ersten Ordnung eine N\"{a}herung daf\"{u}r liefern. Der Unterschied 
zwischen der kinematischen Vorstellung der Bewegung und der von 
uns jetzt beabsichtigten Sch\"{a}tzung eines Mittelwertes f\"{u}r 
die zur\"{u}ckzulegende Strecke besteht also lediglich darin, dass 
wir, statt eines bestimmten Verlaufes anzunehmen, die Wahrscheinlichkeit 
daf\"{u}r berechnen wollen, dass nach n \textit{voneinander unabh\"{a}ngigen} 
Schritten der Abstand von der Anfangslage einen bestimmten Wert 
betr\"{a}gt. Es ist ersichtlich, dass die (klassischen) wahrscheinlichkeitstheoretischen 
Betrachtungen erst nach Preisgabe der Vorstellung, laut welcher 
die vorliegende Anordnung der Schritte mit der Entwicklung von 
sinnvollen Ereignissen verbunden ist, zur Geltung treten k\"{o}nnen.
Dem gem\"{a}{\ss} kann die Bewegung \textit{einer} Zelle weder als passives
Gesteuertwerden noch als aktives Streben gedeutet werden.\\

Sinnvolle bzw. kennzeichnende Ereignisse treten auch
bei mechanischer Reduktion nicht gut ein. Im
Bereich der Mechanik fordert man zwar aus
logischen Gr\"{u}nden, dass jede Ver\"{a}nderung der
Bewegung eines bestimmten, vorgefassten K\"{o}rpers einer
Ursache bed\"{u}rfen soll. Sind die Ursachen von Bestand, 
dann stimmt der Zeitverlauf mit dem aus der Bahn feststellbaren
Werdegang dieses K\"{o}rpers \"{u}berein. Aber die dynamische
Zeit erhellt aus keinem innigen Zusammenhang der Geschehnisse,
sondern sie ist aus der (linearen) dynamischen Darstellung heraus
erkl\"{a}rt worden.
Dass keine zeitliche Entfaltung der Umst\"{a}nde 
vorliegt, leuchtet auch aus folgendem vielgebrauchtem Verfahren ein. Wenn 
wir die Strecke zwischen zwei Marken anhand eines Variationsprinzips 
bestimmen, enth\"{a}lt die Eulersche Differentialgleichung zum 
Problem eine beliebige Funktion des Zeitparameters. Nehmen wir 
nun nach Galilei an, dass bei freier Bewegung die Zeit proportional 
der durchlaufenen Strecke sei. Dann l\"{o}st die Gleichung der 
Gerade das Extremalproblem im Sinne von Newtons erstem Gesetz.
Die Zeit nach Galilei-Newton ist danach derjenige
Parameter, nach dem die geod\"{a}tische Kurve einen 
linearen Ausdruck erh\"{a}lt. Um Lage und Geschwindigkeit 
eines bestimmten Massenpunktes festzusetzen, 
m\"{u}ssen die Anfangsbedingungen dazu genommen
werden, weil sie sich aus der Bahngleichung nicht ergeben.
Sobald nun die Ursachen der Bewegung von den bew\"{a}hrten
Kr\"{a}ften fernab stehen, und die Mechanik nur die Rolle eines Modells
erf\"{u}llt, stellt die Bahn keinen Werdegang mehr dar.
Dessenungeachtet sind von der Mathematik her die 
Anfangsbedingungen genauso willk\"{u}rlich wie eher.
Und ohne diese \textit{jederzeit} frei w\"{a}hlbaren Angaben kennzeichnet man
nicht den Zeitverlauf eines Teilchens, sondern nur die beste
aller m\"{o}glichen Bahnen zwischen zwei Marken.\\

Da nun unserer Meinung nach die mechanische Deutung der Chemotaxis weder
der Steuerung der Diffusion noch der Nachahmung eines Erinnerungsver-m\"{o}gens
dienen kann, setzen wir bis auf weiteres f\"{u}r ein (stichprobenweise) Einzeller:\\
$x_{m}$ die m.te Verschiebung\\
$f_{m}={\pm}1$ eine stochastische Funktion\\
$v_{m}=\alpha\cdot f_{m}\cdot x_m$ die m.te Geschwindigkeit\\
$\Delta t=\Delta t_1=\Delta t_2=\dots=\Delta t_m$ die gemeinsame Zeitspanne.

Das Glied m+1 lautet dann:

\begin{equation}
x_{m+1}=x_m+\nu_m\cdot\Delta t_m
\end{equation}

Anhand der eben gesetzten Definitionen k\"{o}nnen wir die (7) als:

\begin{equation}
x_{m+1}=x_m+\alpha\cdot f_m\cdot x_m\cdot\Delta t
\end{equation}

oder als

\begin{equation}
{x_{m+1}\over x_m}=1+\alpha\cdot f_m\cdot\Delta t
\end{equation}

umschreiben. F\"{u}r m = 1, 2, 3,..., n-1 lautet sie explizite:

\begin{eqnarray}
{x_1\over x_0}&=&1+\alpha\cdot\Delta t\cdot f_0\qquad m=0\nonumber\\
{x_2\over x_1}&=&1+\alpha\cdot\Delta t\cdot f_1\qquad m=1\nonumber\\
&\vdots&\\
{x_{n-1}\over x_{n-2}}&=&1+\alpha\cdot\Delta t\cdot f_{n-2}\qquad m=n-2\nonumber\\
{x_n\over x_{n-1}}&=&1+\alpha\cdot\Delta t\cdot f_{n-1}\qquad m=n-1\nonumber
\end{eqnarray}

Wenn wir in Reih und Glied beiderseits multiplizieren, bilden 
wir das Produkt:

\begin{equation}
{\prod^{m=n-1}_{m=0}{x_{m+1}\over x_m}}=\prod^{m=n-1}_{m=0}{\left (1+\alpha\cdot\Delta t\cdot f_m\right)}
\end{equation}

oder:

\begin{equation}
{{x_1\over x_0}\cdot{x_2\over x_1}\cdot{x_3\over x_2}\dots{x_n\over x_{n-1}}}={
\left(1+\alpha\cdot\Delta t\cdot f_0\right)\dots
\left(1+\alpha\cdot\Delta t\cdot f_{n-1}\right)
}
\end{equation}

das als

\begin{equation}
{x_n\over x_0}={
\prod^{m=n-1}_{m=0}{\left(1+\alpha\cdot\Delta t\cdot f_m\right)}
}
\end{equation}

ausgedr\"{u}ckt werden kann, wobei f$_{m}$ k mal den Wert +1,
und folglich n-k mal den Wert n annimmt, und also x$_{n}$ n verschiedene
Werte hat. Auf die Reihenfolge der Schritte kommt es, wie 
gesagt, nicht an.\\

Was die Statistik \"{u}ber n Zellen betrifft, wollen wir folgendes 
angeben. F\"{u}r k = 0 besteht eine einzige L\"{o}sung, weil f$_{m}$ 
= -1 f\"{u}r jedes m = 1, 2,...,n.\\
F\"{u}r k = 1 sind die L\"{o}sungen n, da man in die (13) f$_{m}$ = 
1 f\"{u}r m = 1, 2,....,n der Reihe nach einzusetzen, und es jedes 
Mal mit den n-1 \"{u}brigen Faktoren, in denen f$_{m}$ den Wert --1 
beibeh\"{a}lt, zu multiplizieren hat.\\
Man kann die kombinatorische Rechnung anwenden, um das allgemeine 
Resultat als

\begin{equation}
x_n = \cases{
x_0\cdot(1-\alpha\cdot\Delta t)^n&${n \choose 0} = 1$\cr
x_0\cdot(1-\alpha\cdot\Delta t)^{n-1}\cdot(1+\alpha\cdot\Delta t)&${n \choose 1} = n$\cr
\qquad\vdots&\cr
x_0\cdot(1-\alpha\cdot\Delta t)^{n-k}\cdot(1+\alpha\cdot\Delta t)^k&${n \choose k} = {n!\over {(n-k)!k!}}$\cr
\qquad\vdots&\cr
x_0\cdot(1+\alpha\cdot\Delta t)^n&${n \choose n}=1$\cr
}
\end{equation}

auszudr\"{u}cken.\\
Aus
${n \choose 0} + {n \choose 1} + \cdots + {n \choose n} =2^{n} $
ermittelt man zum Schluss die Wahrscheinlichkeiten P$_{n}$ als

\begin{equation}
x_n = \cases{
x_0\cdot(1-\alpha\cdot\Delta t)^n&$P_0={1\over 2^n}$\cr
\qquad\vdots&\cr
x_0\cdot(1-\alpha\cdot\Delta t)^{n-k}\cdot(1+\alpha\cdot\Delta t)^k&$P_k={1\over 2^n}\cdot{n!\over {(n-k)!k!}}$\cr
\qquad\vdots&\cr
x_0\cdot(1+\alpha\cdot\Delta t)^n&$P_n={1\over 2^n}$\cr
}
\end{equation}

Ist n gerade, so gibt es ein einziges H\"{a}ufigkeitsmaximum, und 
es betr\"{a}gt:

\begin{equation}
x_n={x_0(1-\alpha\cdot\Delta t)^{n/2}(1+\alpha\cdot\Delta t)^{n/2}}=
{x_0\left\{1-\alpha^2\cdot(\Delta t)^2\right\}^{n/2}}
\end{equation}

Ist dagegen n ungerade, dann ergeben sich zwei gleichwahrscheinliche 
Maxima:

\begin{eqnarray}
x'_n&=&{x_0(1-\alpha\cdot\Delta t)^{(n+1)/2}(1+\alpha\cdot\Delta t)^{(n-1)/2}}=\nonumber\\
&=&{x_0\left(1-\alpha^2\cdot(\Delta t)^2\right)^{(n-1)/2}(1-\alpha\cdot\Delta t)}\nonumber\\
\\
x''_n&=&{x_0(1+\alpha\cdot\Delta t)^{(n+1)/2}(1-\alpha\cdot\Delta t)^{(n-1)/2}}=\nonumber\\
&=&{x_0\left(1-\alpha^2\cdot(\Delta t)^2\right)^{(n-1)/2}(1+\alpha\cdot\Delta t)}\nonumber
\end{eqnarray}

L\"{a}sst man n unbeschr\"{a}nkt wachsen und wird
$\alpha ^{2} \left( \Delta t\right) ^{2} <1$
 gew\"{a}hlt, so strebt diese Folge gegen Null.\\
Da man mit ,,Chemotaxis`` das Wandern von Zellen in Richtung 
eines chemischen Reizes oder fort von ihm bezeichnet, und wir 
Verfahren zur Berechnung vom Mittelwert der Entfernung von dem 
Reiz gebracht haben, erscheint uns unser Ansatz angemessen.\\
Nun m\"{o}chten wir unseren Fall durch die Festlegung einiger Ver\"{a}nderlichen 
erg\"{a}nzen: Zuerst haben wir die Stelle, aus der die chemische 
Anregung geht, auf Null zu setzen. Weiter verlangen wir, dass 
eine fingierte Reizdichte auf solche Weise zu- oder abnimmt, 
dass die Zellengeschwindigkeit linear von der Entfernung abh\"{a}ngt. $\alpha\Delta t$ 
ist ein zus\"{a}tzlicher Parameter. Die Bedingung, dass dessen 
Quadrat kleiner als Eins bleibt, schr\"{a}nkt die zul\"{a}ssigen 
Zellenverschiebungen auf Werte ein, welche kleiner als die jeweilige 
Entfernung vom Ursprung sind.

\section{
Zusammenfassung}

Absicht dieser Schrift ist an einem Beispiel nachzuweisen, dass 
man keiner Vorstellung eines besonderen Mechanismus bedarf, um 
die Chemotaxis zu \textit{beschrei-ben}. Die Modellierung l\"{a}sst sich 
n\"{a}mlich auf enzymatische Reaktionen, deren Anwendungen im Bereich 
der Erreger-Hemmstoffgleichungen liegen, zur\"{u}ckf\"{u}hren. Das bedeutet 
aber, dass man auf die Verwicklung mit komplexen integrierten 
Mechanismen verzichten kann, was bez\"{u}glich Einzeller von Bedeutung 
ist.\\
Die hier gebrachten Modelle k\"{o}nnten anhand von Experimenten
vielleicht weiterentwickelt und erweitert 
werden. Hier begn\"{u}gen wir uns mit dem einfachen Schluss, 
dass ein chemischer Verz\"{o}gerer der Bewegung eine positive (d.h. 
gegen den Ursprung), ein Beschleuniger dagegen eine negative 
Chemotaxis verursacht.

\section*{
Anmerkung}

Das ist meine Wiedergabe einer w\"{a}hrend meines Doktorats auf Englisch geschriebenen Arbeit \cite{viggiano1986}. Auf der ArXiv-Datenbank erscheint sie, knappe f\"{u}nf Jahre nach dem verfr\"{u}hten Tod 
meines Doktorvaters, Herrn Prof. Dr. G. Viggiano.\\

\end{document}